\documentclass[prd,aps,showpacs,twocolumn,nofootinbib]{revtex4}

\usepackage{graphicx}

\begin{document}

\title{Partially ionizing the universe by decaying particles}

\author{S. Kasuya$^a$, M. Kawasaki$^b$, and Naoshi Sugiyama$^c$}

\affiliation{
$^a$ Helsinki Institute of Physics, P. O. Box 64,
     FIN-00014, University of Helsinki, Finland\\
$^b$ Research Center for the Early Universe, 
     University of Tokyo, Tokyo 113-0033, Japan\\
$^c$ Division of Theoretical Astrophysics, 
     National Astronomical Observatory Japan, 
     Tokyo 181-8588, Japan}

\date{September 16, 2003}

\begin{abstract}
We show that UV photons produced by decaying particles can partially
reionize the universe and explain the large optical depth observed by
Wilkinson Microwave Anisotropy Probe. Together with UV fluxes from
early formed stars and quasars, it is possible that the universe is
fully ionized at $z \lesssim  6$ and partially ionized at 
$z \gtrsim 6$ as observed by Sloan Digital Sky Survey for large
parameter space of the decaying particle. This scenario will be
discriminated by future observations, especially by the EE
polarization power spectrum of cosmic microwave background radiation. 
\end{abstract}

\pacs{98.80.-k
\hspace{35mm} HIP-2003-48/TH, astro-ph/0309434}

\maketitle

\setcounter{footnote}{1}
\renewcommand{\thefootnote}{\fnsymbol{footnote}}

\section{Introduction}

It was known for many years that the universe, which once became
almost neutral after recombination, reionized before $z \sim 5$ by
using Gunn-Peterson (GP) test of high redshift quasars.  It was not
until recently, however, that direct evidences to determine the
reionization epoch were found. 

First evidence of the reionization epoch was obtained by Sloan Digital
Sky Survey (SDSS) \cite{SDSS}. 
The spectra of quasars observed by SDSS team revealed that there existed
neutral hydrogen at the redshift $z \simeq 6$ by GP test, which
implies that we have started to see the completion of reionization, 
although the amount of neutral hydrogen does not have to be very large
to explain the observed GP trough. 

Recent Wilkinson Microwave Anisotropy Probe (WMAP) observation 
provided second important evidence of reionization \cite{WMAP}. WMAP
found that the optical depth due to the reionization is 
$\tau_{op} = 0.17 \pm 0.04$, which may correspond to the reionization
epoch $z \sim 20$. 

Therefore, it is observationally required that the reionization
process was very efficient at around $z \sim 20$, but the process
did not complete until $z \sim 6$. The conventional view of the
reionization process is that the universe is reionized by UV fluxes
from early formed stars and quasars. Many calculations have shown
that the universe can be reionized at $z \sim 6 - 10$~\cite{Loeb}. 
Thus, it is possible to explain the full reionization at $z=6$
observed by SDSS but not what WMAP observed. Recent
studies~\cite{Ciardi,FK03} showed that the reionization epoch can be
earlier assuming the non-standard initial mass function (IMF) or
unrealistically large UV fluxes from galaxies. However, the UV fluxes
from stars and quasars increase as time and soon exceed the critical
value to fully reionize the universe. Therefore it is expected that
the universe achieved the full reionization much earlier than 
$z\sim 6$ in this scenario. Therefore we need to consider really
extreme and perhaps unrealistic assumptions about IMF, escaping UV
fraction from galaxies and so on, to have the reionization process
with two stages; for example, see \cite{Astro}.

Alternative to along this line, one may also consider other sources
for ionizing hydrogens and heliums. It could be decaying particles,
but it has been considered that enduring photon emission from particle
decay disables the picture of GP trough confirmed by SDSS. In this
article, however, we will show that the optical depth $\sim 0.17$ can
be explained naturally by decaying particles with rather large
parameter space of mass and lifetime; in some cases, the lifetime
could be longer than the age of the universe. We follow the
evolution of hydrogen, helium, and electron temperature including UV 
sources of quasars, stars, and decaying particles. Here we simply
assume the particles decay into two photons with energy half of the
particle mass. Notice that Hansen and Haiman \cite{HH} tried to
explain the optical depth of $\tau_{op} \approx 0.17$ by a particle of
mass $\sim 200$ MeV and the lifetime  $\sim 4 \times 10^{15}$ sec,
which is $z\approx 20$ reported by WMAP team \cite{WMAP}. They
considered the decaying particle as some sterile neutrino which decay
into electrons (and positrons) in order to achieve half partially
reionized era before $z \sim 6$. Here we will see that photons emitted
through particle decay keeps much smaller partially ionization before
$z \sim 6$, and explains rather large optical depth.

\section{UV sources for ionization}
 
We assume that the particle $\phi$ emits two photons with
monochromatic energy of half mass of that particle, i.e.,
$E_{\gamma}=m_{\phi}/2$. The number density of $\phi$-particle is
written as 
\begin{equation}
    n_{\phi} = n_{\phi}(0) (1+z)^3 e^{-\frac{t}{\tau_{\phi}}},
\end{equation}
where $\tau_{\phi}$ is the lifetime of $\phi$-particle. If 
$E_{\gamma} > 13.6$ eV, those photons emitted can ionize hydrogen
atoms. Then one can write a source term for the decaying particle as
\begin{equation}
    \left(\frac{dn_{\gamma}}{dt}\right)_{dp} 
            = \frac{n_{\phi}(t)}{\tau_{\phi}}.
\end{equation}
In order to calculate how many photons are emitted, the abundance,
mass, and the lifetime of the particle should be fixed. We parameterize
them as free parameters by $\Omega_{\phi}$, $E_{\gamma}$, and
$\tau_{\phi}$. We will see below the allowed region in these parameter
space. Notice that we restrict ourselves only for 
$E_{\gamma} \le 10^5$ eV. This is because $E_{\gamma} \gtrsim$ MeV
case involves e$^+$e$^-$ pair production, and it is too complicated. 
Moreover, $E_{\gamma} \gtrsim O(10)$ MeV may destroy light elements
such as D, T, $^3$He, and $^4$He, spoiling successful big bang
nucleosynthesis. 

Other than decaying particles, we also include usually considered
contributions to reionizing photons from quasars and stars. UV photons
emitted from early formed stars are indeed responsible for achieving 
full ionization at $z\sim 6$. UV photons from quasars may keep the
ionization fraction unity, but this contribution is very small, and
may not be needed for that purpose (Of course, quasars exist and there
are contributions from them). We follow the argument of Fukugita and
Kawasaki \cite{FK94,FK03} to follow the thermal history from $z > 10^3$
including reionization epoch, calculating the ionization fraction of
hydrogen and helium, the photon spectrum, and the electron temperature
(see also \cite{FK90,FK93} for treatment of high energy photons from
decaying particle).

Some of the difference from Ref.~\cite{FK94,FK03} are adoption
of decreasing star formation rate for smaller collapsed objects.
In Refs.\cite{FK94,FK03}  it was assumed that the collapsed objects
which satisfy the cooling condition become star-forming galaxies and
a constant fraction $f$ of baryons go into stars.  In the present
paper, we introduce the star-forming efficiency $f_{star}$ which
depends on the mass of the objects as
\begin{equation}
     f_{star} =\left\{ 
       \begin{array}{cl}
           f \left(\frac{M}{3\times 10^{12}M_{\odot} }\right)^{0.8} & 
           (M < 3\times 10^{12}M_{\odot}), \\[2mm]
           f & (M \ge 3\times 10^{12}M_{\odot}),
       \end{array}\right. 
\end{equation}                    
where $M$ is the mass of the collapsed object.  With $f_{star}$ the 
mass function of galaxies obeys the form of an empirical Schecheter 
function at $z=0$.  Since we expect that the UV fluxes from stars
reionize the universe at relatively low redshift ($z\simeq 6$),  we do
not have to take extreme values for model parameters such as the UV
escape fraction  $F_{UV}$ and $f$. We take $F_{UV} = 0.04$\footnote{
We take $F_{UV} = 0.005 - 0.01$ for longer lifetime than the age of
the universe for $E_{\gamma}=15$ eV.}
and $f=1$ which give the metalicity $Z \approx 0.01$. 

In order to estimate the UV contribution from quasars  we use the
luminosity function in Refs.\cite{Boyle} and \cite{Fan}, from which we
obtain the UV emissivity ( 1/s/cm$^3$) at 4400\AA as
\begin{equation}
    \left(\frac{dn_{\gamma}}{dt}\right)_{QSO} =\left\{ 
      \begin{array}{ll} 
          5\times 10^{-23} 10^{-0.47(z-2)}h &  (z > 2), \\[2mm]
          5\times 10^{-23}h \left(\frac{1+z}{3}\right)^{3.5} 
          & (z \le 2),
      \end{array}\right.
 \end{equation}                    
where $h$ is the Hubble constant in units of 100km/s/Mpc. We also
assume a power low spectrum for quasars,
\begin{equation}
      \left(\frac{d^2n_{\gamma}}{dtdE_{\gamma}}\right)_{QSO}
     \propto \left\{\begin{array}{ll}
           E_{\gamma}^{-0.7} & (E_{\gamma} < 10.2~{\rm eV}),\\[2mm]
           E_{\gamma}^{-1.4} & (E_{\gamma} \ge10.2~{\rm eV}).
       \end{array}\right.
\end{equation}

We set the cosmological parameters following WMAP-only \cite{WMAP}:  
$H_0=72$~km/s/Mpc, $\Omega_m=0.29$, $\Omega_{\Lambda}=1-\Omega_m$, 
$\Omega_b=0.047$, and $\sigma_8=0.9$. Apparently, we are dealing with
$\Lambda$CDM universe, and the decaying particle is assume to be a
part of cold dark matter,\footnote{
Particles with very small mass such as $\lesssim 10^5$ eV can be
regarded as cold dark matter in such a way that a scalar condensate is
corehently oscillating.}
but could be hot dark matter with negligible contribution to the total
density of the universe.

\section{Ionization history}

\begin{figure}[!t]
\includegraphics[width=80mm]{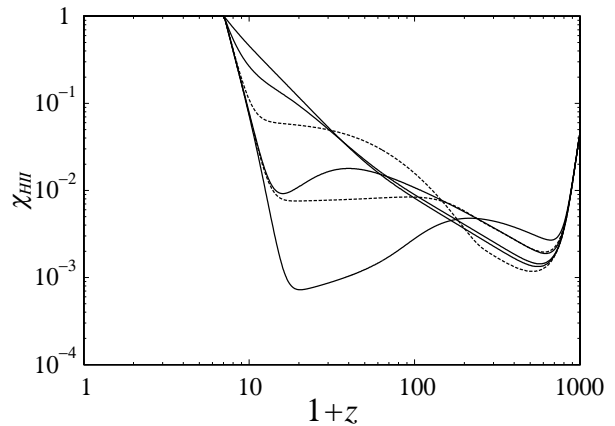}
\caption{\label{HII}
Ionization histories of hydrogen (HII). We plot for
$E_{\gamma}=15$~eV for $\tau_{\phi}=10^{14}$, $10^{15}$, $10^{16}$,
and $10^{18}$ sec in solid lines from the bottom to the top (at 
$z \sim 20$), while lower and upper dashed lines denote for
$E_{\gamma}=10^5$ eV for $\tau_{\phi}=10^{14}$ and $10^{15}$ sec,
respectively.}
\end{figure}

We follow the evolution of the fraction of ionized hydrogen and
helium, in addition to electron temperature. Figure \ref{HII} shows
ionization history of hydrogen for $E_{\gamma}=15$ and $10^5$ eV for
various lifetime $\tau_{\phi}$. In each case, the abundance of
$\phi$-particles is determined in such a way that the optical depth
reads $\tau_{op}\simeq 0.17$, which we define by
\begin{equation}
    \tau_{op} = \int_{0}^{\infty} dz \sigma_T 
    \left(\frac{dt}{dz}\right)[n_{HII}-n_{HII}\big|_{sr}],
\end{equation}
where $\sigma_T$ is the Thomson cross section and $n_{HII}\big|_{sr}$
is the number density for standard recombination. We subtract this
term in order to estimate only the effect of reionization.

\begin{figure}[!t]
\includegraphics[width=80mm]{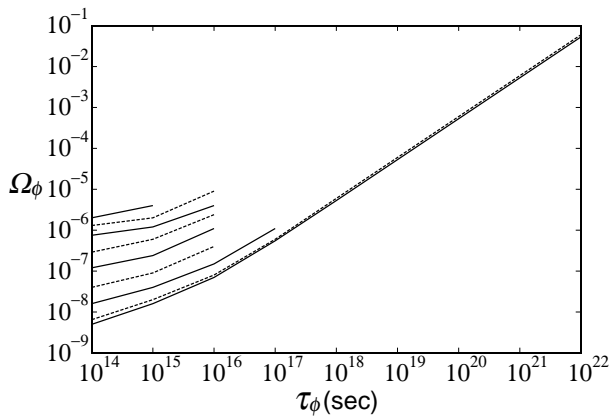}
\caption{\label{Eall}
Allowed region leading to $\tau_{op}\simeq 0.17$. Lines show for
$E_{\gamma}=15$, 30, $10^2$, $3\times 10^2$, $10^3$, $3\times 10^3$,
$10^4$, $3\times 10^4$, and $10^5$ eV from the bottom to the
top. We exclude those regions where the photon flux exceeds the cosmic
X-ray background: typically, $E_{\gamma} \gtrsim 10^2$ eV and
$\tau_{\phi} \gtrsim 10^{17}$~sec.}
\end{figure}

In cases for $E_{\gamma} \lesssim 30$ eV, the lifetime of
$\phi$-particle is allowed for many orders of magnitude. For longer
lifetime than the age of the universe, it is the value of
$n_{\phi}/\tau_{\phi}$ that concerns with the amount of photon
emission, so ionization histories of these cases are identical to that
of $\tau=10^{18}$ sec shown in Fig.\ref{HII}. Notice that the upper
limit of the lifetime is around $4 \times 10^{22}$ sec, when 
$\phi$-particles is (cold) dark matter, i.e., 
$\Omega_{\phi} \simeq \Omega_m$. For shorter lifetime, the abundance
should be much less than that contributes to the energy density of the
universe. See Fig.~\ref{Eall}. 

On the other hand, if larger energy of photons are emitted, we have to
be careful about cosmic X-ray backgrounds. We adopt the following
simple formulas to compare with the calculated photon flux \cite{KY}: 
\begin{eqnarray}
    F_{\gamma,obs} & \simeq & 
    8 \left(\frac{E_{\gamma}}{{\rm keV}}\right)^{-0.4}
    (0.2 {\rm keV} \lesssim E_{\gamma} \lesssim 25 {\rm keV}), \\
    & \simeq & 
    380 \left(\frac{E_{\gamma}}{{\rm keV}}\right)^{-1.6} 
    (25 {\rm keV} \lesssim E_{\gamma} \lesssim 350 {\rm keV}), \\
    & \simeq &
    2 \left(\frac{E_{\gamma}}{{\rm keV}}\right)^{-0.7}
    (350 {\rm keV} \lesssim E_{\gamma} \lesssim 2 {\rm MeV}), 
\end{eqnarray}
where $F_{\gamma,obs}$ is measured in units of 
(cm$^2$ sr sec)$^{-1}$. 

We find that long lifetime is not allowed, because the emitted photon
flux exceeds the observed diffuse photon background. Thus, the
lifetime is rather restricted in the narrow range, and should
typically be $\tau_{\phi} \lesssim 10^{16}$ sec. This constraint is
taken into account in Fig.~\ref{Eall}, thus the parameter space is
restricted for $E_{\gamma} \gtrsim 10^2$ eV. Hydrogen ionizing
histories for $E_{\gamma}=10^5$ eV are shown in Fig.~\ref{HII} as an
example. They look very similar to those for $E_{\gamma}=15-30$ eV. In
this case, even $\tau_{\phi}=10^{16}$ sec seems to be excluded as is
shown in Fig.~\ref{Eall}. 

\begin{figure}[t]
\includegraphics[width=80mm]{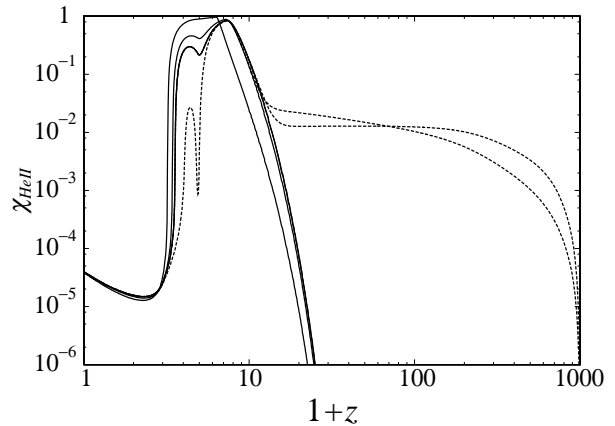}
\caption{\label{HeII}
Ionization histories of helium (HeII). We plot for
$E_{\gamma}=15$~eV for $\tau_{\phi}=10^{14}$, $10^{15}$, $10^{16}$,
and $10^{18}$ sec in solid lines from right to left (near
$z \sim 20$), while lower and upper dashed lines denote for
$E_{\gamma}=10^5$ eV for $\tau_{\phi}=10^{14}$ and $10^{15}$ sec,
respectively.}
\end{figure}

\begin{figure}[!t]
\includegraphics[width=80mm]{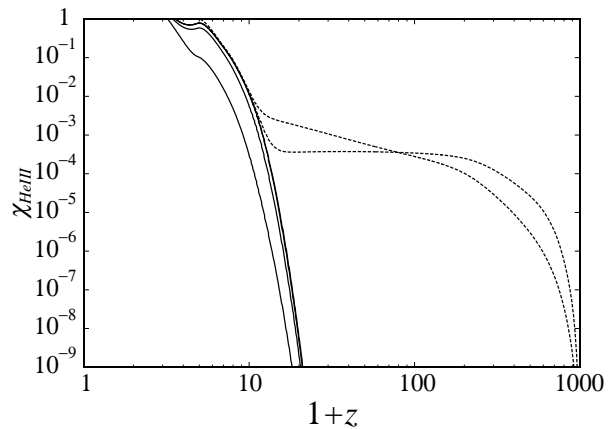}
\caption{\label{HeIII}
Ionization histories of helium (HeIII). We plot for
$E_{\gamma}=15$~eV for $\tau_{\phi}=10^{14}$, $10^{15}$, $10^{16}$,
and $10^{18}$ sec in solid lines from right to left (near 
$z \sim 20$), while lower and upper dashed lines denote for
$E_{\gamma}=10^5$ eV for $\tau_{\phi}=10^{14}$ and $10^{15}$ sec,
respectively.}
\end{figure}

Evolutions of ionization fraction of hydrogen look more or less
similar in any emitted photon energy for the same lifetime. In order 
to distinguish them, we need to look at the ionization history of
helium. Since larger photon energy can excite helium atoms, one can
see the qualitative difference from low energy emitted photons. In
Figs.~\ref{HeII} and \ref{HeIII}, ionization histories of HeII and
HeIII are shown, respectively. We only plotted for $E_{\gamma}=15$ and
$10^5$ eV cases as in Fig.~\ref{HII}.

It is trivial to say that rather high ionization fraction begins at
high redshift in large $E_{\gamma}$ cases. Therefore, although it
will be very difficult to observe ionization fraction of helium at 
$z \gtrsim 20$, it might be one of the tools for (dis)proving the
particle decay scenario.

\section{Cosmic microwave background spectrum}
\begin{figure}[!t]
\includegraphics[width=80mm]{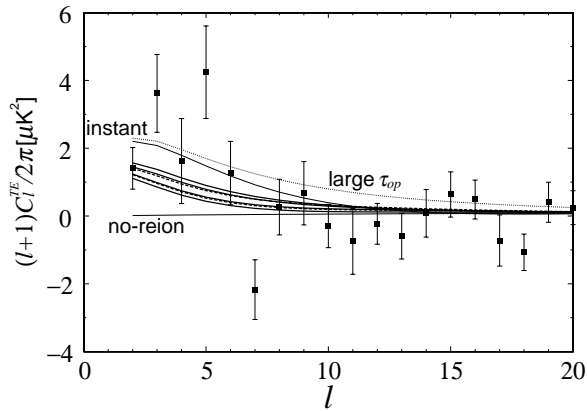}
\caption{\label{te}
TE spectrum for various ionization histories. 
We plot for $E_{\gamma}=15$~eV for $\tau_{\phi}=10^{14}$,
$10^{15}$, $10^{16}$, and $10^{18}$ sec from the bottom to the top in
thick solid lines, while lower and upper thick dashed lines denote for
$E_{\gamma}=10^5$ eV for $\tau_{\phi}=10^{14}$ and $10^{15}$ sec,
respectively. We plot no and instantaneous reionization cases in thin
solid lines. Higher optical depth case is also shown in dotted line 
($E_{\gamma}=15$ eV, $\tau_{\phi}=10^{16}$ sec, and 
$\tau_{op} \simeq 0.43$). In addition, WMAP data is placed.}  
\end{figure}

The evolutions of ionization fraction of hydrogen and helium
look very different from those instantaneous reionization history
usually considered in the literatures, although we obtain the optical
depth of $\sim 0.17$. So it is interesting to know what we can see, or
more precisely, whether we can distinguish them, in the cosmic
microwave background (CMB) power spectrum. Figure \ref{te} shows the
cross correlation spectra between temperature and E-mode
polarization. Our models seem to be less power in the small $\ell$
region than the model with instantaneous reionization. This is because
particle decay takes place for longer period, so the power spreads to
larger $\ell$ for fixed value of the optical depth. From the TE
spectrum, it is not apparent to distinguish between these models
within these error bars. Even difference between the instantaneous
reionization and ionization by decaying particles can hardly
seen. Moreover, notice that larger optical depth $\tau_{op} \simeq
0.43$ may account for the observational results of the TE spectrum, 
too. Of course, it is known that larger optical depth makes the first
acoustic peak of the TT spectrum lower. However it can be adjusted by
introducing larger spectral index, so-called $\tau_{op}-n$ degeneracy.

\begin{figure}[!t]
\includegraphics[width=80mm]{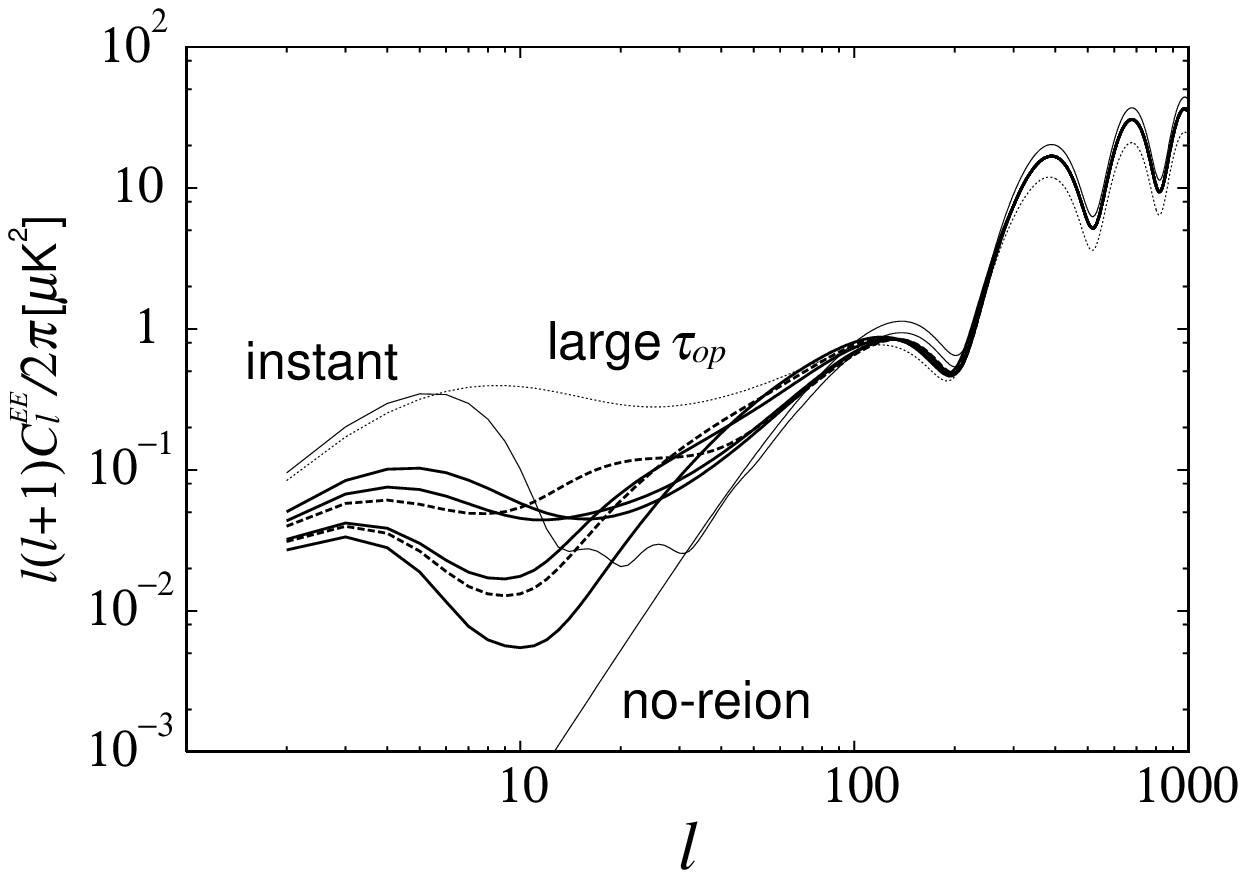}
\caption{\label{ee}
EE spectrum for various ionization histories. 
We plot for $E_{\gamma}=15$~eV for $\tau_{\phi}=10^{14}$,
$10^{15}$, $10^{16}$, and $10^{18}$ sec from the bottom to the top (at
$z\sim 10$) in thick solid lines, while lower and upper thick dashed
lines denote for $E_{\gamma}=10^5$ eV for $\tau_{\phi}=10^{14}$ and
$10^{15}$ sec, respectively. We plot no and instantaneous reionization
casea in thin solid lines. Higher optical depth case is also shown in
dotted line ($E_{\gamma}=15$ eV, $\tau_{\phi}=10^{16}$ sec, and 
$\tau_{op} \simeq 0.43$). }  
\end{figure}

Unlike TE spectra, however, EE spectra seem to serve us a much
powerful tool for discriminating among different ionization
histories,\footnote{
Even in more conventional cases, differences of EE spectra seem 
to have ability to distinguish among different ionization 
histories~\cite{Holder,HuHolder}.} 
as shown in Fig.~\ref{ee}. Peculiar feature of enduring photon
emission from particle decays is larger power around 
$\ell=$ (a few)$\times 10$. In addition, there is different feature
between long and short lifetime: for long lifetime, the spectrum is
rather flat, while there is a dip at $\ell \sim 10$ for short
lifetime. Thus, EE spectrum hoped to be seen by Planck satellite or
even WMAP should (dis)prove the photons from decaying particles in the
near future.
 
To end this section, we comment on the Compton $y_c$-parameter. We
also follow its evolution, and obtained that the final value is
(several)$\times 10^{-7}$ in all the parameters we choose, which is
well below the current observational limit \cite{COBE}.

\section{Conclusion and discussion}

We have studied comprehensively the ionization history of the universe
by photons emitted from decaying particles, together with the effects
of stars, quasars, and shown that the decaying particle can explain
rather large optical depth observed by WMAP without conflicting other
astrophysical and cosmological constraints. One of the amazing fact is
that the lifetime of order $10^{15}$ sec (in other words, $z\sim 20$)
is not necessarily required for successful scenario, and larger
parameter space is allowed. Since the early star formation ejecting UV
photons may be difficult to lead large optical depth without adjusting
parameters beyond their natural values, this alternative is at least
as good as them. 

For $E_{\gamma}=10^2 - 10^5$ eV, larger lifetime is prohibited,
because the emitted flux exceeds the observed diffuse photon
background, and $\tau_{\phi}\simeq 10^{14}-10^{16}$ sec is only
allowed in these cases. On the other hand, if some bump is observed in
the photon spectrum, it could be a great clue for decaying particle
scenario. 

Longer lifetime is also allowed for $E_{\gamma}\lesssim 30$ eV. In the
limiting case, the decaying particle can be the dark matter, provided
that its lifetime is around $4 \times 10^{22}$ sec. Notice that our
situation is different from Ref.~\cite{Sciama}, in which they
investigated for decaying neutrino dark matter and thus using hot dark
matter scenario. In this case, ionization fraction becomes unity at 
$z \sim 20-30$, which conflicts with SDSS results. In addition, EURD
observation may falsify decaying particle scenario with 
$E_{\gamma} \sim 15$ eV, but it needs careful look, since their 
conclusion will change if the abundance of galactic dark matter near
the Sun is one third of their applied value \cite{EURD}. 

Observation of CMB spectrum is very powerful. Although instantaneous
reionization suggests optical depth $\tau_{op}\sim 0.17$, higher
optical depth may be favored for the observed TE spectrum by WMAP if
the decaying particle scenario is correct, where slow reionization
takes place. It is somewhat difficult to discriminate from each other
only by TE spectrum. However, the differences in the EE spectrum is
tremendous so we hope that even the lifetime of the decaying particle
will be measured by Planck.

Finally, we comment on what can be a candidate for this decaying
particle. In the context of no-scale supergravity, scalar particles
remain massless at the tree level. They may acquire their masses at
one-loop order. Among them, saxion, the scalar partner of axion in
supersymmetric theory, might be the best candidate. If we consider the
hadronic axion model, the lifetime is estimated as \cite{GY} 
\begin{equation}
    \tau_s \simeq 10^{15} 
      \left(\frac{F_{PQ}}{10^{15}{\rm GeV}}\right)^2
      \left(\frac{m_s}{100 {\rm eV}}\right)^{-1} {\rm sec},
\end{equation}
where $F_{PQ}$ is the axion decay constant. It perfectly fits into the
allowed region shown in Fig.~\ref{Eall}.

\section*{Acknowledgments}
NS is supported by Japanese  Grant-in-Aid for Science Research Fund of
the Ministry of Education, No.14340290.



\end{document}